\PassOptionsToPackage{dvipsnames}{xcolor}

\documentclass[sigconf,natbib,9pt]{acmart}

\usepackage{algorithm}
\usepackage{algpseudocode}
\usepackage{graphicx}
\usepackage{subfigure} 
\usepackage{amsmath}
\usepackage{multirow}
\usepackage{hyperref}
\usepackage[capitalise]{cleveref}
\usepackage{xcolor}
\usepackage{natbib}
\usepackage{paralist}
\usepackage{beramono}
\usepackage{listings}
\usepackage[dvipsnames]{xcolor}
\usepackage{enumitem}
\usepackage{setspace}

\definecolor{dkgreen}{rgb}{0,0.6,0}
\definecolor{gray}{rgb}{0.5,0.5,0.5}
\definecolor{mauve}{rgb}{0.58,0,0.82}
\definecolor{codebackground}{rgb}{0.95, 0.95, 0.95}
\definecolor{codeborder}{rgb}{0.85, 0.85, 0.85}
\definecolor{codeinline}{rgb}{0.1, 0.1, 0.1}

\lstdefinestyle{scala}{
	frame=tb,
	language=scala,
	aboveskip=3mm,
	belowskip=3mm,
	showstringspaces=false,
	columns=flexible,
	basicstyle={\Small\ttfamily},
	numbers=none,
	numberstyle=\tiny\color{gray},
	keywordstyle=\color{blue},
	commentstyle=\color{dkgreen},
	stringstyle=\color{mauve},
	frame=single,
	rulecolor=\color{codeborder},
	backgroundcolor=\color{codebackground},
	breaklines=true,
	breakatwhitespace=true,
	tabsize=3,
	xleftmargin=2mm,
	xrightmargin=2mm,
}

\newcommand{\OurLDA}{APS-LDA}
\newcommand{\OurPS}{Glint}
\newcommand{\CompFramework}[1]{cluster computing framework#1}

\newcommand{\Pipeline}[1]{data processing pipeline#1}

\newcommand\footnoteDataset{\footnote{ \url{http://cake.da.inf.ethz.ch/clueweb-topicmodels/}}}

\title{Computing Web-scale Topic Models using an Asynchronous Parameter Server}

\author{Rolf Jagerman}
\orcid{}
\affiliation{%
\institution{University of Amsterdam}
\city{Amsterdam}
\country{The Netherlands}
}
\email{rolf.jagerman@uva.nl}

\author{Carsten Eickhoff}
\orcid{}
\affiliation{%
\institution{ETH Z\"urich}
\city{Z\"urich}
\country{Switzerland}
}
\email{carsten.eickhoff@inf.ethz.ch}

\author{Maarten de Rijke}
\orcid{0000-0002-1086-0202}
\affiliation{%
\institution{University of Amsterdam}
\city{Amsterdam}
\country{The Netherlands}
}
\email{derijke@uva.nl}

\begin{document}
	
\begin{abstract} 
Topic models such as Latent Dirichlet Allocation (LDA) have been widely used in information retrieval for tasks ranging from smoothing and feedback methods to tools for exploratory search and discovery.
However, classical methods for inferring topic models do not scale up to the massive size of today's publicly available Web-scale data sets. The state-of-the-art approaches rely on custom strategies, implementations and hardware to facilitate their asynchronous, commu\-nication-intensive workloads.

We present \OurLDA{}, which integrates state-of-the-art topic modeling with \CompFramework{s} such as Spark using a novel \emph{asynchronous} parameter server. Advantages of this integration include convenient usage of existing \Pipeline{s} and eliminating the need for disk writes as data can be kept in memory from start to finish.
Our goal is not to outperform highly customized implementations, but to propose a general high-performance topic modeling framework that can easily be used in today's \Pipeline{s}. We compare \OurLDA{} to the existing Spark LDA implementations and show that our system can, on a 480-core cluster, process up to 135$\times$ more data and 10$\times$ more topics without sacrificing model quality.
\end{abstract}


\copyrightyear{2017} 
\acmYear{2017} 
\setcopyright{acmlicensed}
\acmConference{SIGIR '17}{}{August 07-11, 2017, Shinjuku, Tokyo, Japan}\acmPrice{15.00}\acmDOI{http://dx.doi.org/10.1145/3077136.3084135}
\acmISBN{978-1-4503-5022-8/17/08}


\maketitle


\section{Introduction}
\label{sec:introduction}

Probabilistic topic models are a useful tool for discovering a set of latent themes that underlie a text corpus~\cite{hofmann1999probabilistic,blei2003latent}. Each topic is represented as a multinomial probability distribution over a set of words, giving high probability to words that co-occur frequently and small probability to those that do not.

Recent information retrieval applications often require very large-scale topic modeling to boost their performance~\citep{yuan2015lightlda}, where many thousands of topics are learned from terabyte-sized corpora. Classical inference algorithms for topic models do not scale well to very large data sets. This is unfortunate because, like many other machine learning methods, topic models would benefit from a large amount of training data. 

When trying to compute a topic model on a Web-scale data set in a distributed setting, we are confronted with a major challenge:

\begin{center}\emph{How do individual machines keep their model synchronized?}\end{center}

\noindent%
To address this issue, various distributed approaches to LDA have been proposed. The state-of-the-art approaches rely on custom strategies, implementations and hardware to facilitate their asynchronous, commu\-nication-intensive workloads~\cite{chen2015warplda,yu2015scalable,yuan2015lightlda}. These highly customized implementations are difficult to use in practice because they are not easily integrated in today's data processing pipelines.

\begin{figure}
	\includegraphics[width=\columnwidth]{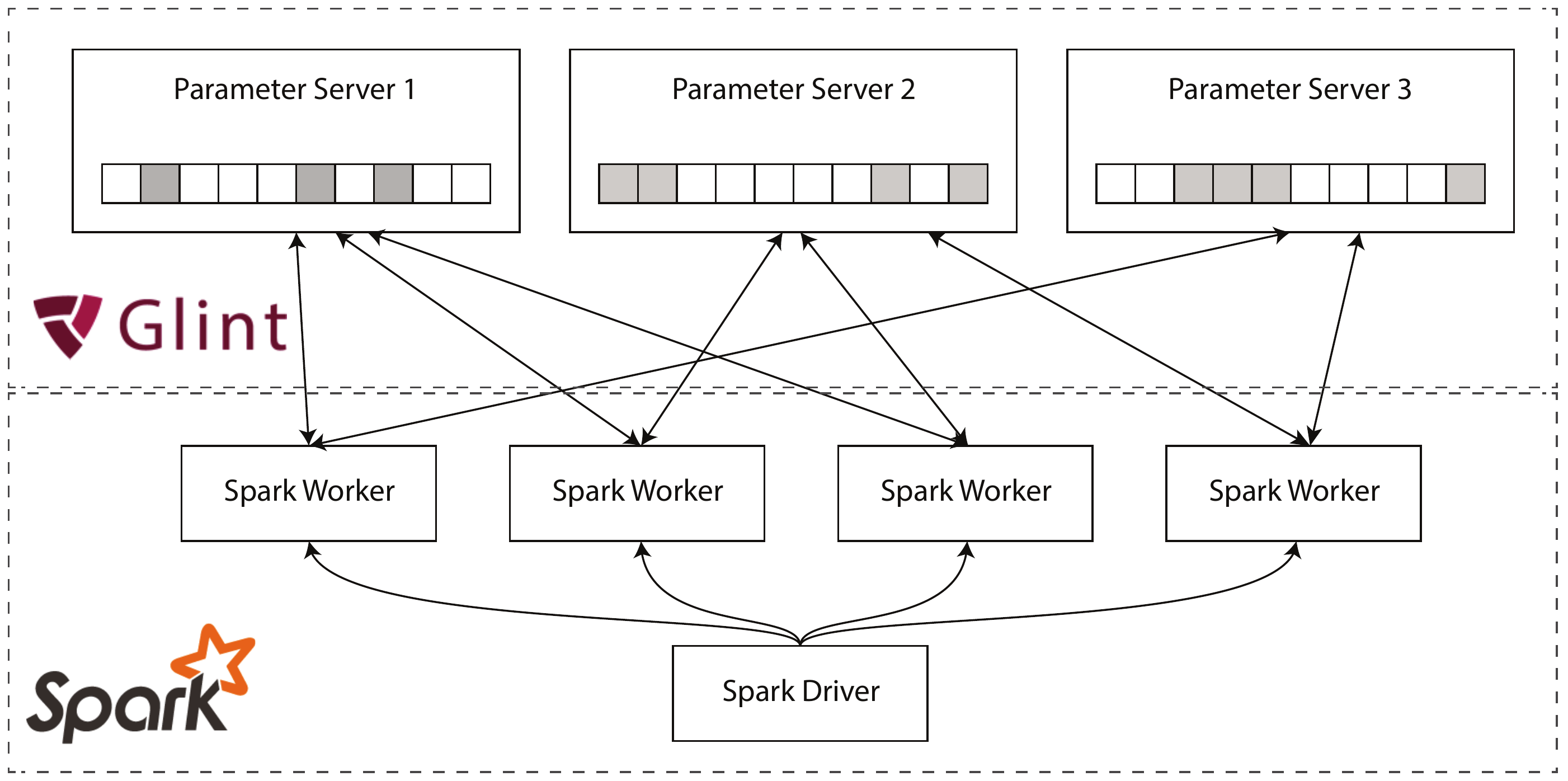}
	\caption{High-level overview of the Glint parameter server architecture and its interaction with Spark. The parameter servers provide a distributed and concurrently accessed parameter space for the model being learned.}
	\label{fig:overview}
\end{figure}

We propose \OurLDA{}, a distributed version of LDA that builds on a widely used \CompFramework{}, Spark~\citep{zaharia2012resilient}. The advantages of integrating model training with existing \CompFramework{s} include convenient usage of existing data-processing pipelines and eliminating the need for intermediate disk writes since data can be kept in memory from start to finish~\citep{moritz-sparknet-2016}. However, Spark is bound to the typical map-reduce programming paradigm. Common inference algorithms for LDA, such as collapsed Gibbs sampling, are not easily implemented in such a paradigm because they rely on a large mutable parameter space that is updated concurrently. We address this by adopting the para\-meter server model~\citep{li2014scaling}, which provides a distributed and concurrently accessed para\-meter space for the model being learned (see \cref{fig:overview}).

\section{Distributed LDA}
\label{sec:distributedlda}
We present \OurLDA, our distributed version of LDA, which builds on the LightLDA algorithm~\cite{yuan2015lightlda}; it uses an asynchronous version of the parameter server, as we will detail in Section~\ref{sec:architecture}.

\subsection{LightLDA}
LightLDA performs a procedure known as collapsed Gibbs sampling, which is  a Markov Chain Monte-Carlo type algorithm that assigns a topic $z \in \{1, \ldots, K\}$ to every token in the corpus. It then repeatedly re-samples the topic assignments $z$. The LightLDA algorithm provides an elegant  method for re-sampling the topic assignments in $\mathcal{O}(1)$ time by using a Metropolis-Hastings sampler. This is important because sampling billions of tokens is computationally infeasible if every sampling step would use $\mathcal{O}(K)$ operations, where $K$ is a potentially large number of topics.


To re-sample the topic assignments $z$, the algorithm needs to keep track of the statistics $n_{k}$, $n_{wk}$ and $n_{dk}$:

\begin{itemize}[nosep,leftmargin=14pt]
	\item $n_k$: Number of times any word was assigned topic $k$
	\item $n_{wk}$: Number of times word $w$ was assigned topic $k$
	\item $n_{dk}$: Number of times a token in document $d$ was assigned topic $k$
\end{itemize}

\noindent%
It is clear that the document-topic counts $n_{dk}$ are document-specific and thus local to the data and need not be shared across machines. However, the word-topic counts $n_{wk}$ and topic counts $n_k$ are global and require sharing. The parameter server provides a shared interface to these values in the form of a distributed matrix storing $n_{wk}$, and a distributed vector storing $n_k$.

\subsection{\OurLDA{}: A Re-design of LightLDA}
Despite its attractive properties, LightLDA has an important shortcoming. It uses a stale-synchronous parameter server in which push requests are batched together and sent once when the algorithm finishes processing its current partition of the data. This architecture uses a fixed network thread and may cause a stale model, where individual machines are unable to see updates from other machines for several iterations.

In contrast, our approach sends push requests \emph{asynchronously} during the compute stage. These more frequent but smaller updates have a number of essential advantages:
\begin{enumerate}[nosep,leftmargin=14pt]
	\item It decreases the staleness of the model while it is computing. With our approach it is possible to see updates from other machines within the same iteration over the data, something that is not possible with the standard parameter server.
	\item It makes mitigating network failure easier as small messages can be resent more efficiently.
	\item It enables the algorithm to take advantage of more dynamic threading mechanisms such as fork-join pools and cached thread pools \cite{welc2005safe}.
\end{enumerate}

\noindent%
The move from such a fixed threaded design to a fully asynchronous one requires a re-design of LightLDA. Algorithm~\ref{alg:lightlda} describes the \OurLDA{} method. At the start of each iteration, the algorithm performs a synchronous pull on each processor $p$ to get access to the global topic counts $n_k$. It then iterates over the vocabulary terms, and asynchronously pulls the word-topic counts $n_{wk}$ (line~\ref{lst:asynclda:syncpull}). These asynchronous requests call back the {\sc Resample} procedure when they complete. The {\sc Resample} procedure (line~\ref{ourlda:11}) starts by computing an alias table on the available word-topic counts $n_{wk}$. This alias table is a datastructure that can sample from a categorical probability distribution in amortized $\mathcal{O}(1)$ time. The algorithm then iterates over the local partition of the data $\mathcal{D}_p$ where it resamples every (token, topic) pair using LightLDA's $\mathcal{O}(1)$ Metropolis-Hastings sampler, which requires the earlier mentioned alias table. Changes to the topic counts are pushed asynchronously to the parameter server while it is computing (\cref{lst:asynclda:asyncpush1,lst:asynclda:asyncpush2,lst:asynclda:asyncpush3,lst:asynclda:asyncpush4})

Note that all of our push requests either increment or decrement the counters $n_{wk}$ and $n_k$. The parameter server exploits this fact by aggregating these updates via addition, which is both commutative and associative. This eliminates the need for complex locking schemes that are typical in key-value storage systems. Instead, the updates can be safely aggregated through an atomic integer structure that is easy to implement.

\begin{algorithm}[t]
	\begin{algorithmic}[1]
		\State {$\mathcal{P} \leftarrow $ Set of processors},\\
		{$\mathcal{D} \leftarrow $ Collection of documents},\\
		{$\mathcal{V} \leftarrow $ Set of vocabulary terms}
		\Statex\vspace{-0.2cm}
		\For{$p \in \mathcal{P}$ in parallel}
		\State $\mathcal{D}_p \subseteq \mathcal{D}$
		\State $n_k \leftarrow $ {\sc SyncPull}($\{n_k \mid k = 1 \dots K \}$) \label{lst:asynclda:syncpull}
		\For{$w \in \mathcal{V}$}
		\State{\bf on} {\sc AsyncPull}($\{n_{wk} \mid k = 1 \dots K \}$)
		\State\hspace{4.5mm}{\bf call} {\sc Resample($\mathcal{D}_p, n_{wk}, n_k$)} \label{lst:asynclda:asyncpull}
		\EndFor
		\EndFor
		\Statex\vspace{-0.2cm}
		\Procedure {Resample}{$\mathcal{D}_p, n_{wk}, n_k$}\label{ourlda:11}
		\State{$a \leftarrow $ AliasTable($n_{wk}$)}
		\For{$(w, z_{\text{old}}) \in d \in \mathcal{D}_p$}
		\State{$z_{\text{new}} \leftarrow$ MetropolisHastingsSampler($a, d, w, z_{\text{old}}, n_k, n_{wk}$)}\label{lst:asynclda:lightlda}
		\State{\sc AsyncPush}($\{n_{wk} \leftarrow n_{wk} + 1\}$) for $k = z_{\text{new}}$ \label{lst:asynclda:asyncpush1}
		\State{\sc AsyncPush}($		\{n_k \leftarrow n_k + 1\}$) for $k = z_{\text{new}}$ \label{lst:asynclda:asyncpush2}
		\State{\sc AsyncPush}($\{n_{wk} \leftarrow n_{wk} - 1\}$) for $k = z_{\text{old}}$\label{lst:asynclda:asyncpush3}
		\State{\sc AsyncPush}($\{n_k \leftarrow n_k - 1\}$) for $k = z_{\text{old}}$\label{lst:asynclda:asyncpush4}
		\EndFor
		\EndProcedure
	\end{algorithmic}
	\caption{\OurLDA: Asynchronous Parameter Server LDA.}
	\label{alg:lightlda}
\end{algorithm}

In the next section, we will discuss the asynchronous parameter server that makes the implementation of this algorithm possible.

\section{Parameter Server Architecture}
\label{sec:architecture}

The traditional parameter server architecture~\cite{li2013parameter} is a complete machine learning framework that couples task scheduling, a distributed (key, value) store for the parameters and user-defined functions that can be executed on workers and servers. As a result, there is considerable complexity in the design, setup and implementation of a working parameter server, making it difficult to use in practice.

We present \OurPS{},\footnote{\url{https://github.com/rjagerman/glint/}} an open-source asynchronous parameter server implementation. Our implementation is easily integrated with the \CompFramework{} Spark, which allows us to leverage Spark features such as DAG-based task scheduling, straggler mitigation and fault tolerance. This integration is realized by decoupling the components of the traditional parameter server architecture and removing the dependency on task scheduling. This is accomplished by simplifying the parameter server interface to a set of two operations:

\begin{enumerate}[nosep,leftmargin=14pt]
	\item \textbf{Asynchronously `Pull' data from the servers.}\\This will query parts of the matrix or vector.
	\item \textbf{Asynchronously `Push' data to the servers.}\\This will update parts of the matrix or vector.
\end{enumerate}

\noindent%
The goal of our parameter server implementation is to store a large distributed matrix and provide a user with fast queries and updates to this matrix. In order to achieve this, it will partition and distribute the matrix to multiple machines. Each machine only stores a subset of rows. Algorithms interact with the matrix through the \textit{pull} and \textit{push} operations, unaware of the physical location of the data.

\subsection{Pull action}
Whenever an algorithm wants to retrieve entries from the matrix it will call the \textit{pull} method. This method triggers an asynchronous pull request with a specific set of row and column indices that should be retrieved. The request is split up into smaller requests based on the partitioning of the matrix such that there will be at most one request per parameter server.

Low-level network communication provides an `at-most-once' guarantee on message delivery. This is problematic because it is impossible to know whether a message sent to a parameter server is lost or just takes a long time to compute. However, since pull requests do not modify the state of the parameter server, we can safely retry the request multiple times until a successful response is received. To prevent flooding the parameter server with too many requests, we use an exponential back-off timeout mechanism. Whenever a request times out, the timeout for the next request is increased exponentially. If after a specified number of retries there is still no response, we consider the pull operation failed.

\subsection{Push action}
In contrast to pull requests, a \textit{push} request will modify the state on the parameter servers. This means we cannot na\"{i}vely resend requests on timeout because if we were to accidentally process a push request twice it would result in a wrong state on the parameter server. We created a hand-shaking protocol to guarantee `exactly-once' delivery on push requests.\footnote{\url{https://github.com/rjagerman/glint/blob/master/src/main/scala/glint/models/client/async/PushFSM.scala}} The protocol first attempts to obtain a unique transaction $id$ for the push request. Data is transmitted together with the transaction $id$, allowing the protocol to later acknowledge receipt of the data. A timeout and retry mechanism is only used for messages that are guaranteed not to affect the state of the parameter server. The result is that pushing data to the parameter servers happens exactly once.

\subsection{LDA implementation}
\begin{figure}
	\centering
	\includegraphics[width=.91\columnwidth]{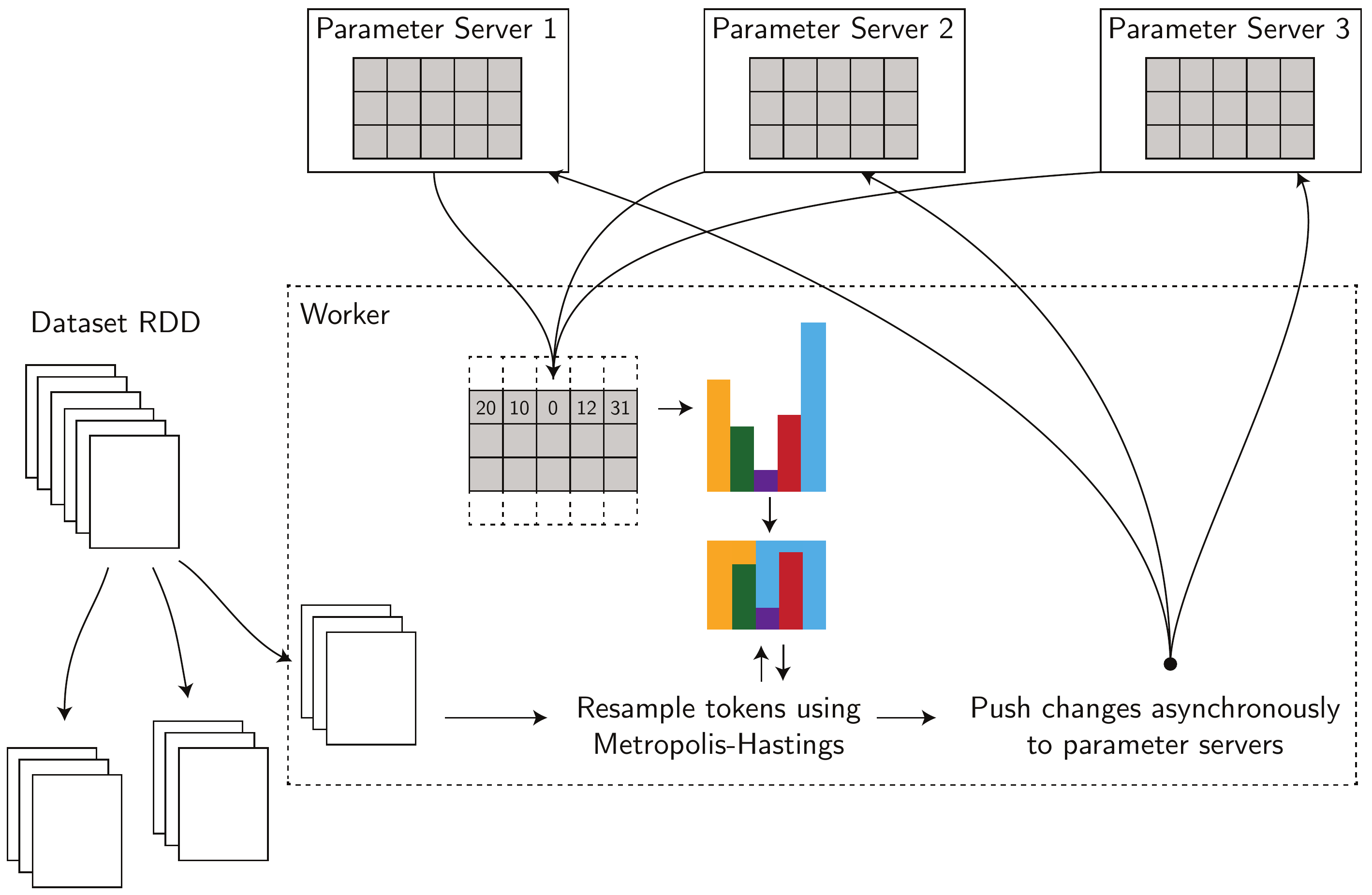}
	\vspace*{-0.5\baselineskip}

	\caption{Overview of the implementation. A dataset is split into different partitions by Spark. Tokens in each partition are resampled by the Metropolis-Hastings algorithm. Updates are pushed asynchronously to the parameter server.}
	\label{fig:archoverview}
\end{figure}

We have implemented the \OurLDA{} algorithm using Spark and the asynchronous parameter server. A general overview of the implementation is provided in \cref{fig:archoverview}. The Spark driver distributes the Resilient Distributed Dataset (RDD) of documents to different workers. Each worker pulls parts of the model from the parameter server and constructs corresponding alias tables. The worker then iterates over its local partition of the data and resamples the tokens using the Metropolis-Hastings algorithm. Updates are pushed asynchronously to the parameter server while the algorithm is running.

\section{Experiments}
\label{sec:experiments}

There is no point in optimizing and scaling inference if the quality of the trained model should suffer. For this reason, we want to validate that the effectiveness of the trained model remains the same. It should be noted that our goal is \textbf{not} to outperform highly customized implementations such as LightLDA. 

Instead, we aim to integrate state-of-the-art topic modeling with Spark such that large topic models can be efficiently computed in modern \Pipeline{s}. To this end, we compare our implementation against existing Spark implementations on the same hardware and configuration. We compare \OurLDA{} to two existing state-of-the-art LDA algorithms provided by Spark's MLLib: The EM algorithm \cite{asuncion2009smoothing} and the online algorithm \cite{hoffman2010online}. We run our experiments on a compute cluster with 30 nodes, with a total of 480 CPU cores and 3.7TB RAM. The nodes are interconnected over 10Gb/s ethernet. The Clue\-Web12~\cite{clueweb12} corpus, a 27-terabyte Web crawl that contains 733 million Web documents, is used as the data set for our experiments.

\begin{figure}[t]
	\centering
		\centering
		\vspace*{-\baselineskip}
		\includegraphics[width=0.81\columnwidth,trim=1cm 0cm 0.3cm 0cm]{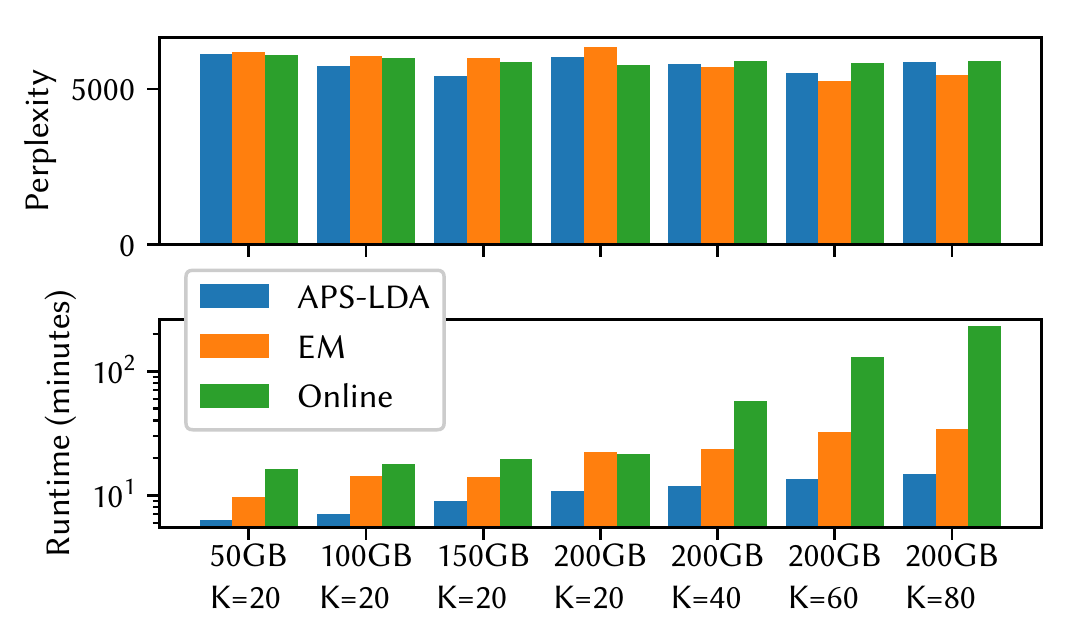}
		
		\vspace*{-\baselineskip}
		\caption{Performance of APS-LDA compared to the EM~\cite{asuncion2009smoothing} and Online~\cite{hoffman2010online} algorithms for different data set sizes (50GB--200GB) and different numbers of topics $K$ (20--80).}
		\label{fig:performance}
\end{figure}

To validate that our methods do not sacrifice the quality of the trained model we will compare the three algorithms on small subsets of ClueWeb12. We vary either the number of topics (20--80) or the size of the data set (50GB--200GB) to measure how the different systems scale with those variables and use perplexity as an indicator for topic model quality. Due to the large size of the data, a hyperparameter sweep is computationally prohibitively expensive and we set the LDA hyperparameters $\alpha = 0.05$ and $\beta = 0.001$ which we found to work well on the ClueWeb12 data set.
We split the data in a 90\% training set and a 10\% test set and measure perplexity on the test set. \cref{fig:performance} shows the results of the experiments. We observe that, barring some variations, the perplexity is roughly equal for all algorithms. However, our implementation has a significantly better runtime. We use a log-scale for the runtime in minutes.

\begin{figure}[t]
	\begin{minipage}{.46\textwidth}
		\centering
		\vspace*{-\baselineskip}
		\includegraphics[width=0.81\columnwidth,trim=0.3cm 0.35cm 1.1cm 1.2cm]{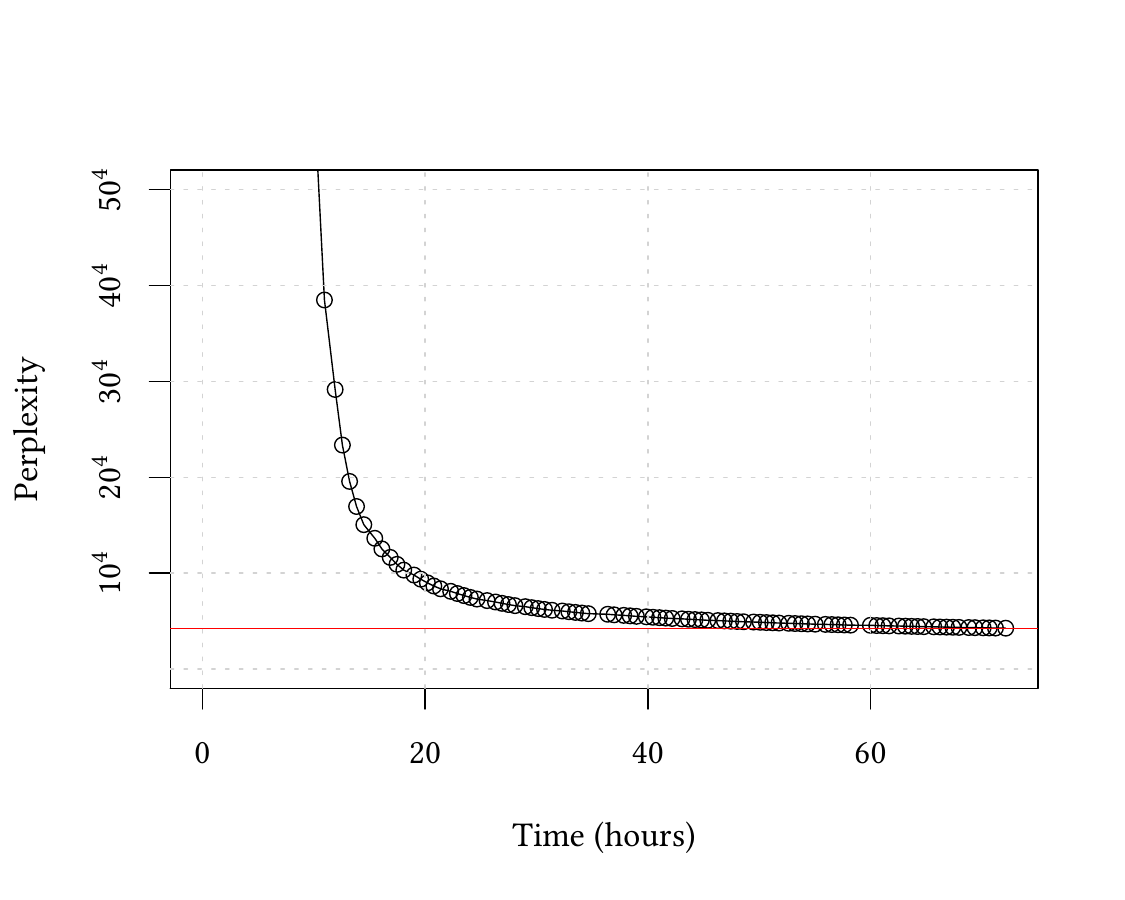}
		
		\vspace*{-\baselineskip}
		\caption{Perplexity of the 1,000-topic LDA model on ClueWeb12.}
		\label{fig:perplexity}
	\end{minipage}
\end{figure}

When attempting to increase the data set size beyond 200GB, the default Spark implementations cause numerous failures due to an increase in runtime and/or shuffle write size. Our implementation is able to effortlessly scale far beyond these limits and compute an LDA model on the full ClueWeb12 data set (27TB) with 1,000 topics in roughly 80 hours (see~\cref{fig:perplexity}). This is an increase of nearly two orders of magnitude, both in terms of dataset size and number of  topics, using identical hardware and configuration. We have made the final 1,000-topic LDA model publicly available in CSV format.\footnoteDataset


\section{Conclusion}
\label{sec:conclusion}

We have presented \OurLDA{}, a distributed method for computing topic models on Web-scale data sets. It uses an asynchronous parameter server that is easily integrated with the \CompFramework{} Spark. We conclude our work by revisiting the challenge that was presented in the introduction:
 
\begin{center}\emph{How do individual machines keep their model synchronized?}\end{center}

\noindent%
 The asynchronous parameter server solves this by providing a distributed and concurrently accessed parameter space for the model being learned. The asynchronous design has several advantages over the traditional parameter server model: it prevents model staleness, makes mitigating network failure easier and enables the system to use more dynamic threading mechanisms.

Our proposed algorithm \OurLDA{}, is a thorough re-design of LightLDA that takes advantage of the asynchronous parameter server model. We have implemented this algorithm and the asynchronous parameter server using Spark, a popular \CompFramework{}. The resulting architecture allows for the computation of topic models that are several orders of magnitude larger, in both dataset size and number of topics, than what was achievable using existing Spark implementations. The code of \OurLDA{} is available as open source (MIT licensed) and we are also sharing a 1,000-topic LDA model trained on ClueWeb 12.

Finally, there are two promising directions for future work:
\begin{inparaenum}
	\item Large-scale information retrieval tasks often require machine learning methods such as factorization machines and deep learning, which are known to benefit from the parameter server architecture~\cite{dean2012large}. By using an asynchronous parameter server, it may be possible to achieve significant speedups.
	\item Our current implementation of the asynchronous parameter server uses a dense representation of the data, due to the garbage collection constraint imposed by the JVM runtime. By implementing sparse representations it is possible to scale even further as this will reduce both memory usage and network communication overhead.
\end{inparaenum}

\medskip
\begin{spacing}{1}
\noindent\small
\textbf{Acknowledgments.}
This research was supported by
Ahold Delhaize,
Amsterdam Data Science,
the Bloomberg Research Grant program,
Criteo,
the Dutch national program COMMIT,
Elsevier,
the European Community's Seventh Framework Programme (FP7/2007-2013) under
grant agreement nr 312827 (VOX-Pol),
the Microsoft Research Ph.D.\ program,
the Netherlands Institute for Sound and Vision,
the Netherlands Organisation for Scientific Research (NWO)
under pro\-ject nrs
612.001.116, 
HOR-11-10, 
CI-14-25, 
652.\-002.\-001, 
612.\-001.\-551, 
652.\-001.\-003, 
and
Yandex.
All content represents the opinion of the authors, which is not necessarily shared or endorsed by their respective employers and/or sponsors.
\end{spacing}

\vspace*{-.1\baselineskip}
\bibliographystyle{abbrvnatnourl}
\bibliography{parameterserver} 


\end{document}